**Title:**

# Direct programming of confined Surface Phonon Polariton Resonators using the plasmonic Phase-Change Material In$_3$SbTe$_2$

*Author(s), and Corresponding Author(s)\**

*Lukas Conrads[+,\*], Luis Schüler[+], Konstantin G. Wirth, Matthias Wuttig, Thomas Taubner[#]*

+ authors contributed equally

**Affiliations**

Institute of Physics (IA), RWTH Aachen University, D-52056 Aachen, Germany.

\* Email: conrads@physik.rwth-aachen.de

# Email: taubner@physik.rwth-aachen.de

**Abstract:** (200 words)

Tailoring light-matter interaction is essential to realize nanophotonic components. It can be achieved with surface phonon polaritons (SPhPs), an excitation of photons coupled with phonons of polar crystals, which also occur in 2d materials such as hexagonal boron nitride or anisotropic crystals. Ultra-confined resonances are observed by restricting the SPhPs to cavities. Phase-change materials (PCMs) enable non-volatile programming of these cavities based on a change in the refractive index. Recently, the new plasmonic PCM In$_3$SbTe$_2$ (IST) was introduced which can be reversibly switched from an amorphous dielectric state to a crystalline metallic one in the entire infrared to realize numerous nanoantenna geometries. However, reconfiguring SPhP resonators to modify the confined polaritons modes remains elusive. Here, we demonstrate direct programming of confined SPhP resonators by phase-switching IST on top of a polar silicon carbide crystal and investigate the strongly confined resonance modes with scanning near-field optical microscopy. Reconfiguring the size of the resonators themselves result in enhanced mode confinements up to a value of $\lambda/35$. Finally, unconventional cavity shapes with complex field patterns are explored as well. This study is a first step towards rapid prototyping of reconfigurable SPhP resonators that can be easily transferred to hyperbolic and anisotropic 2d materials.

**Introduction:**

Tailoring the light-matter interaction at the nanoscale is one of the main goals in nanophotonics. The research area of plasmonics facilitates subdiffractional field confinement by exploiting metallic nanostructures. However, those structures suffer from extensive losses due to the very short lifetime of the excited surface waves called surface plasmon polaritons.[1] This obstacle can be solved by exploiting surface phonon polaritons (SPhPs) occurring in polar crystals facilitating low-loss nanophotonics. The dielectric function of polar crystals offers a narrow region called Reststrahlenband between the longitudinal and transversal optical phonon frequency with a negative value of the real part and a small imaginary part of the dielectric function. SPhPs can be excited at an interface of a dielectric and the polar crystal and comprise of coupled photons with collective oscillations of atomic cores. One of the most prominent polar crystals is silicon carbide (SiC) with the Reststrahlenband ranging from 797 – 973 $cm^{-1}$.[2–7] A large variety of applications has been demonstrated in the past years, for example in thermal management[8][9], metamaterials[10,11], sensing[12,13], or non-linear optics[14].

Unfortunately, the functionality and design wavelength of those devices is fixed after fabrication and cannot easily be altered afterwards.

Dynamic functionalities can be achieved by employing different mechanisms such as the photoexcitation of free charge carriers in SiC[15] or combining SPhPs with active materials. For example, the temperature dependent insulator-to-metal transition of $VO_2$ enables tuning of the wavelength of phonon polaritons[16]. However, local modifications are not possible due to the volatile and temperature induced phase transitions.

Phase-change materials (PCMs) are prime candidates for non-volatile tuning based on a change in the refractive index between two (meta-) stable phases, e.g., the amorphous phase and the crystalline one. Remarkably, both phases show a large difference in the optical properties, caused by a change in chemical bonding. In the amorphous phase, the atoms are covalently bonded, but in the crystalline phase a new bonding type occurs which is called metavalent[17–21]. The reversible phase-change can be induced by locally heating the PCM with electrical or optical pulses. Consequently, those materials have been utilized for reconfigurable polariton optical elements such as lenses[22] and for switchable SPhP resonators[23,24].

Recently, the new plasmonic PCM In$_3$SbTe$_2$ (IST) was introduced[25], which can be reversibly switched from a dielectric amorphous state to a metallic crystalline one in the entire infrared range. In particular, the permittivity of crystalline IST is negative and follows a Drude-like behavior, as shown in **Figure 1A**. In the crystalline phase, the electrons are strongly delocalized and enhance the conductivity of the material. The conductivity is about $10^4$ S/cm and therefore allows the classification as a bad metal. Previously, arbitrary nanoantenna shapes were fabricated via direct laser writing and even a large-area spatial emissivity shaping metasurface was demonstrated[26–29]. However, reconfiguring the resonators to modify the confined polaritons modes and investigating the coupling of metallic IST resonators to polaritons remains elusive.

Here, we demonstrate direct writing of confined SPhP resonators via laser irradiation with the plasmonic PCM IST (see **Figure 1B**) on top of the polar crystal silicon carbide (SiC). Therefore, the PCM is locally heated with a focused laser above the glass transition temperature to crystallize the material. Crystallizing the amorphous IST (aIST) around the targeted resonator shape offers a significant degree of freedom for programming arbitrary shapes of amorphous IST within a crystalline surrounding. First, we investigate freely propagating SPhPs launched by an edge of crystalline IST, followed by confined SPhP modes of circular cavities with scattering-type scanning near-field optical microscopy (s-SNOM). Reducing the diameter of the fabricated cavities results in a spectral shift of the observed resonance modes. Afterwards, the field confinement is determined and stronger confinements for smaller cavities are obtained. Finally, we study arbitrary resonator shapes such as squares and triangles to highlight the vast flexibility of programming different shapes with the plasmonic PCM IST.

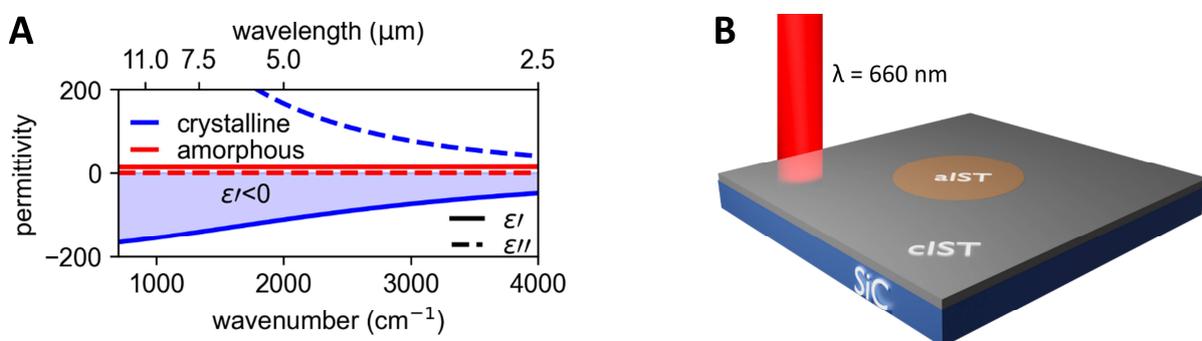

**Figure 1: A)** Dielectric function of amorphous and crystalline IST. Upon crystallization, the PCM changes from an amorphous dielectric to a crystalline metallic state. **B)** The PCM is crystallized via laser irradiation to form amorphous cavities of IST on top of a SiC substrate.

**Propagating Surface Phonon Polaritons:**

First, we investigate freely propagating SPhPs launched by a laser crystallized edge of crystalline IST with s-SNOM (c.f. **Figure 2A**). S-SNOM enables subwavelength resolution achieved by probing the sample with highly localized near-fields of a sharp tip. The near-field amplitude and phase are detected by demodulating the scattered electric field of the oscillating tip illuminated by infrared laser light. Therefore, the 35 nm thin amorphous IST layer on top of the polar crystal SiC is crystallized with precise laser pulses (see **Methods**). The propagating SPhPs are launched at the crystalline edge and propagate along the amorphous IST. **Figure 2B** displays the measured optical amplitude images $s_2$, normalized to crystalline IST, for frequencies varied from 880 to 930 cm$^{-1}$. Bright periodic fringes appear next to the crystalline IST, corresponding to SPhPs. Upon increasing the frequency, the spacing between these fringes increases. The polariton wavelength is determined by extracting line profiles along the measured s-SNOM images (see **Figure 2C**) and applying a fit for a damped oscillating function with an additional linear background (see **Supplementary Note 2** for more details and a zoom-in for better visibility of the fitted function onto the measured data). The measured propagation wavevectors from Figure 2C are compared with theoretical calculations of the dispersion of SPhPs in **Figure 2D**. The green curve represents the calculated slow-guided mode and the brown curve the fast-guided mode of the polaritons excited at the three-layer interface SiC/aIST/air. Both modes occur due to the chosen complex $k_{sp}$ ansatz (see **Supplementary Note 3** for more information).[23] Here, the calculated solution only takes three layers into account by neglecting the capping layer on top of the IST layer. Another solution is obtained by calculating the imaginary part of the reflection coefficient $r_p$ (color-coded map in Figure 2D) with the transfer matrix method (see **Supplementary Note 3**). The maximum of the imaginary part of the reflection coefficient corresponds to the excitation of SPhPs and is close to the calculated slow mode of the polaritons. The experimentally extracted data (red dots) match well with the simulated imaginary part of the reflection coefficient. The measured propagation length of the SPhPs is in the range of 1 µm, allowing for programming confined resonators of similar size.

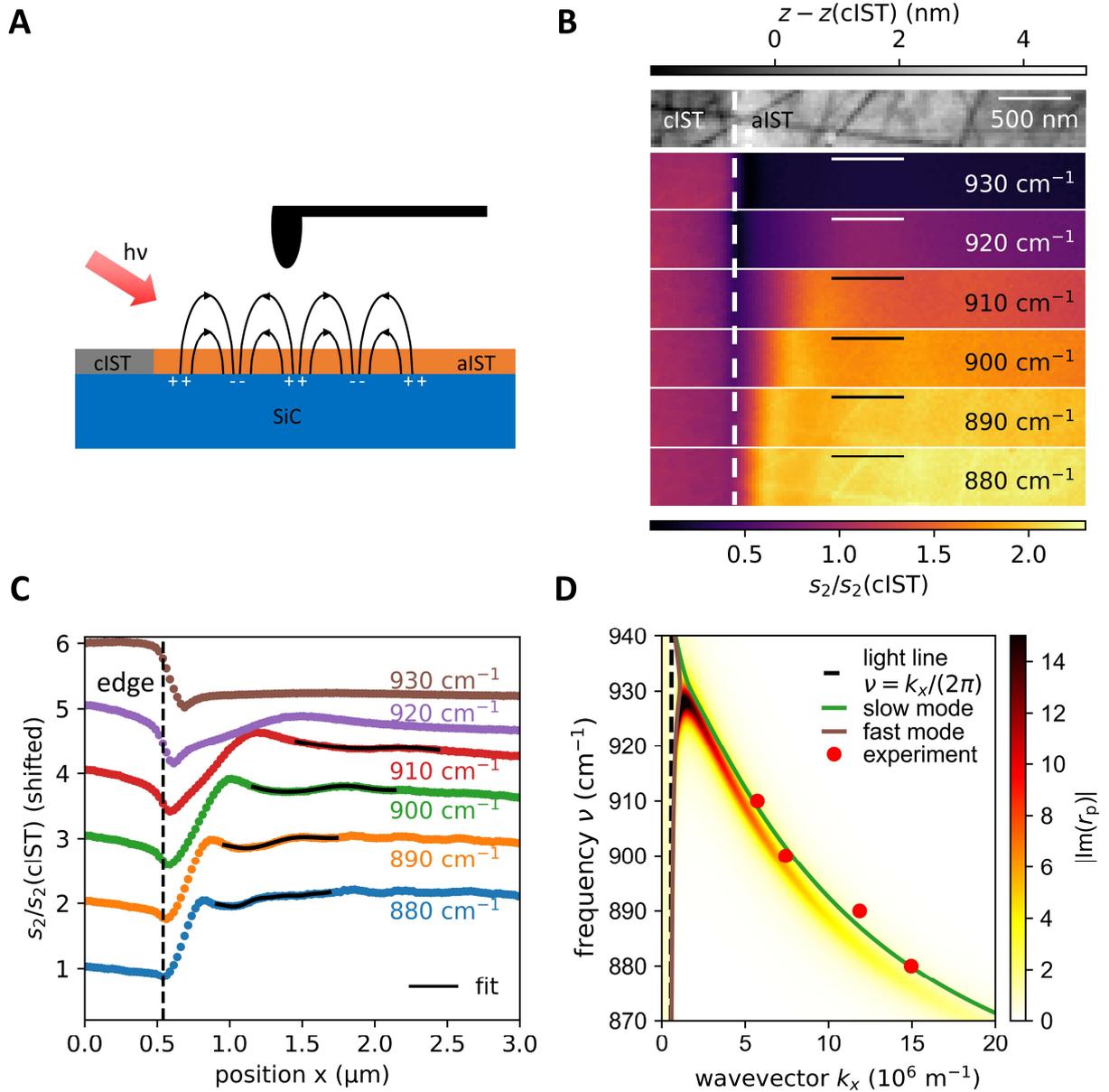

**Figure 2: Propagating SPhPs launched by a cIST edge. A)** Surface phonon polaritons are launched at the laser crystallized edge and are detected with s-SNOM. **B)** Optical near-field amplitude images $s_2$ and the corresponding AFM image (top) of the SPhPs for different excitation frequencies. The images are normalized to the signal of crystalline IST. **C)** Averaged line profiles of the amplitude near-field images $s_2$ taken from **B**. The black line corresponds to a fitted function to determine the wavevector of the polaritons. **D)** Dispersion plot of the SPhP on a SiC substrate covered with amorphous IST and capping. The green curve represents the expected slow mode of the polaritons for a layer stack without capping. A more adequate solution is given by the imaginary part of the reflection coefficient (color-coded) with capping considered. The experimentally obtained values are shown as red dots.

**Confined Surface Phonon Polariton Resonators:**

Next, we optically write a circular SPhP resonator of amorphous IST on SiC with a diameter of $D$ = 3 µm by crystallizing the material around the targeted structure (see **Methods** for more details). The measured s-SNOM near-field amplitude images with additional cross-section profiles and the topography can be seen in **Figure 3A**. The excitation frequency is continuously decreased from 935 cm$^{-1}$ to 875 cm$^{-1}$. The near-field amplitude is normalized to the signal of the surrounding crystalline IST, which does not exhibit any spectral features. A pronounced and confined field enhancement appears at 920 cm$^{-1}$. Decreasing the excitation frequency leads to an evolution of the observed amplitude pattern inside the resonator. In particular, a ring mode structure is visible at 910 cm$^{-1}$. This pattern further transforms for an excitation frequency of 900 cm$^{-1}$ into a ring with enhanced fields in the center. For smaller frequencies, no mode structure can be observed.

Afterwards, we decreased the diameter of the cavity by applying more crystallizing laser pulses to investigate the effect of smaller cavities on the observed resonance modes (see **Figure 3B**). The procedure of creating those resonator shapes with sophisticated spatially overlapping laser pulses is explained in **Supplementary Note 4**. The corresponding s-SNOM amplitude images for the different cavity diameters can be found in **Supplementary Note 5**. The optical near-field amplitude in the center of the cavity for varied frequencies and different cavity diameters is shown in **Figure 3C**. All curves display a maximum in the optical amplitude at the center which is shifted towards smaller frequencies for reduced cavity diameters. In addition, a cut-off is observed for wavenumbers larger than this peak. Theoretically, the resonance frequency of such circular cavities can be calculated with $k_{\mathrm{sp}}(\nu)D + \phi = 2x_n(J_m)$[30], where $k_{\mathrm{sp}}(\nu)$ is the polariton wavevector, $D$ corresponds to the cavity diameter, and $\phi$ and $x_n(J_m)$ denote the reflection phase and the $n$th zero of the Bessel function $J_m$, respectively. We derived a reflection phase of -π which is in accordance with the expectations due to the metallic boundaries of the cavity (see **Supplementary Note 6** for more details).[6,31] With our presented technique of directly writing cavities featuring various resonance modes, programming of antenna arrays is possible as well. These arrays display clear resonances in the far-field spectra and are shown in **Supplementary Note 7**.

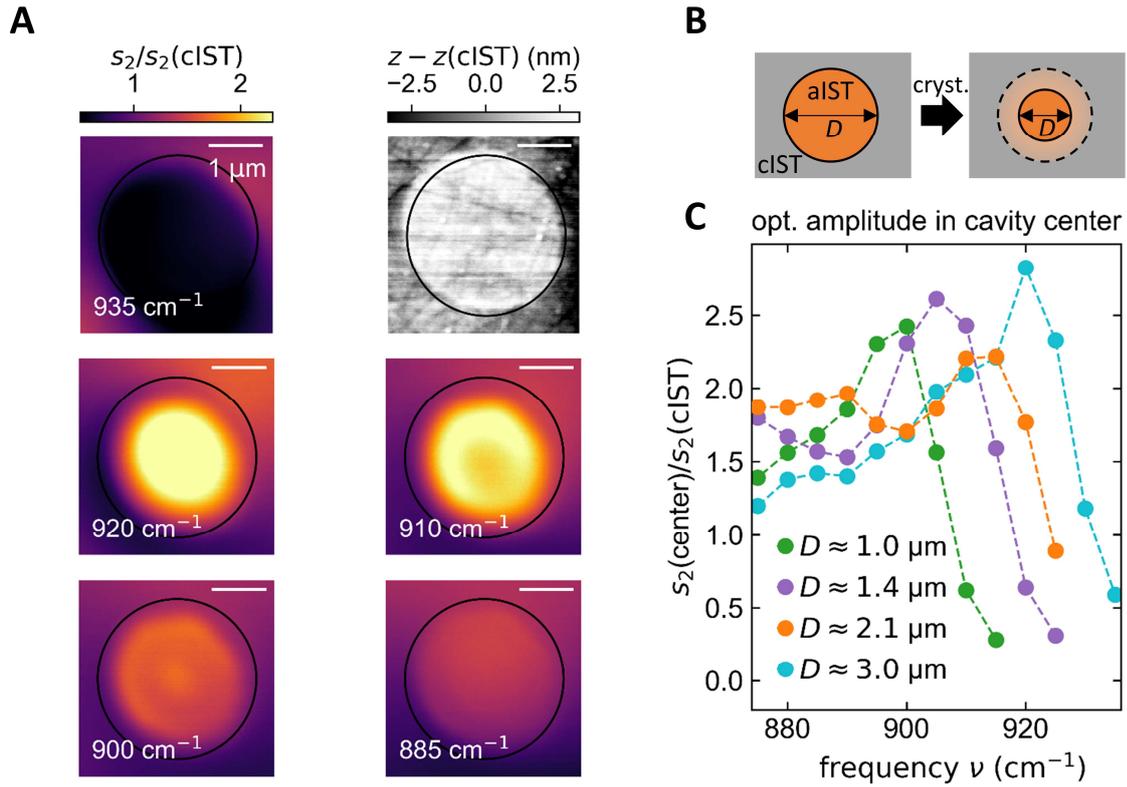

**Figure 3: A)** Normalized near-field amplitude images $s_2$ of a cavity with $D$ = 3 µm for excitation frequencies from 875 cm$^{-1}$ to 935 cm$^{-1}$. Only a few representative s-SNOM images are displayed. The upper right image displays the topography of the amorphous cavity with a height of 2.3 nm. **B)** Adding more crystallizing pulses at the edge of the cavity results in reconfigured cavities with a smaller diameter $D$. **C)** Measured normalized near-field amplitude in the cavity center for different diameters as a function of the excitation frequency. For smaller diameters, the peak shifts towards smaller frequencies.

For further insights of the confined fields inside the cavities, we evaluated the lateral extension of the measured near-field amplitude at the corresponding resonance frequencies for different diameters in **Figure 4**. While for all different cavities, a pronounced maximum is observed at their resonance frequency, the lateral extension varies. Reducing the cavity diameter results in more confined fields. Notice the different scale bars of the s-SNOM images. To quantify the field confinements, the FWHM of a line profile is determined. In the cavity with the largest diameter of $D$ = 3 µm, the confinement is $\lambda_0/5.5$. The confinement can be improved by a factor of 7 by reducing the diameter of the cavity to $D$ = 1 µm, resulting in a field confinement of $\lambda_0/35$. Hence, reconfiguring the size of the cavities themselves allows for tuning the field confinement far below the diffraction limit of light.

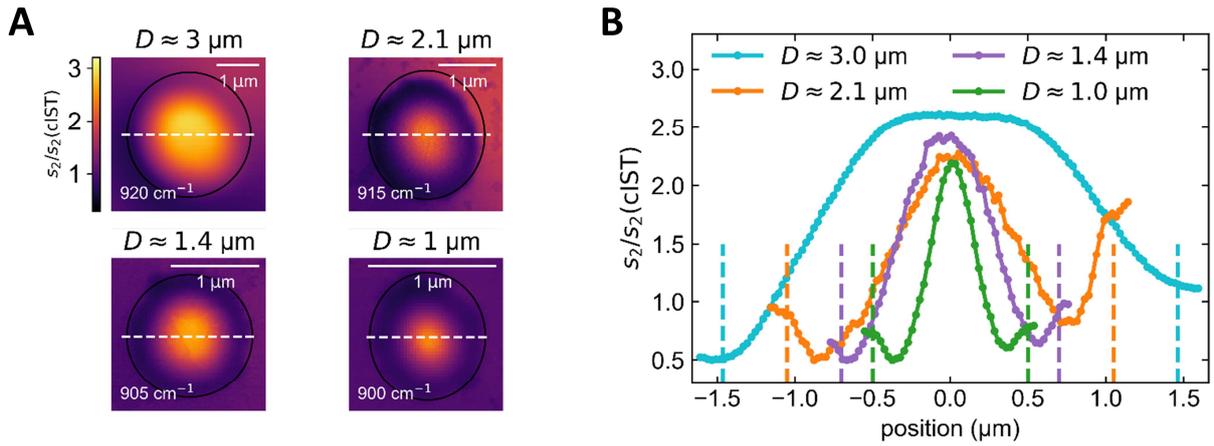

**Figure 4: Mode confinement of circular resonators with varied diameter. A)** Optical near-field amplitude images of the (0,1) mode for different cavity diameters at the corresponding excitation frequencies. **B)** Cross-section profile of the different cavities from A. Reducing the cavity diameter results in more confined fields. For the smallest cavity of 1.0 µm, a field confinement with a diameter of 320 nm is achieved.

Conventional fabrication techniques of SPhP resonators are associated with tremendous effort and precision. Accordingly, we highlight the advantage of direct laser writing targeted geometries by investigating more complex cavity geometries which can be directly written by applying laser pulses. **Figure 5A** displays the light microscope image of a fabricated square resonator with 3 µm edge length. The corresponding measured near-field images at varied frequencies are shown in **Figure 5B**. The near-field amplitude obtained inside the resonator strongly depends on the chosen excitation frequency: Alternating patterns of enhanced and attenuated amplitude fields are shown for increasing excitation frequency. Numerical field simulations of the out-of-plane component of the electric field are displayed in **Figure 5C**. Remarkably, the experimentally obtained patterns are well reproduced, although the simulations exhibit more details because small imperfections of the laser-written resonators disturb the optical near-field image.

Subsequently, we have investigated an exceptional resonator shape of a triangle with edge lengths of 2.7 µm (see light microscope image in **Figure 5D**). The corresponding measured near-field amplitude images are shown in **Figure 5E** for the same excitation frequencies investigated previously. Here, the lateral field extension inside the triangular resonator becomes more confined for increased frequency because the pattern transitions from a higher to a lower mode. The comparison with numerical field simulations (c.f. **Figure 5F**) reveals that the experimentally obtained data only match qualitatively with the simulations. Fabrication imperfections and roughness might add an additional loss channel attenuating the observable pattern.

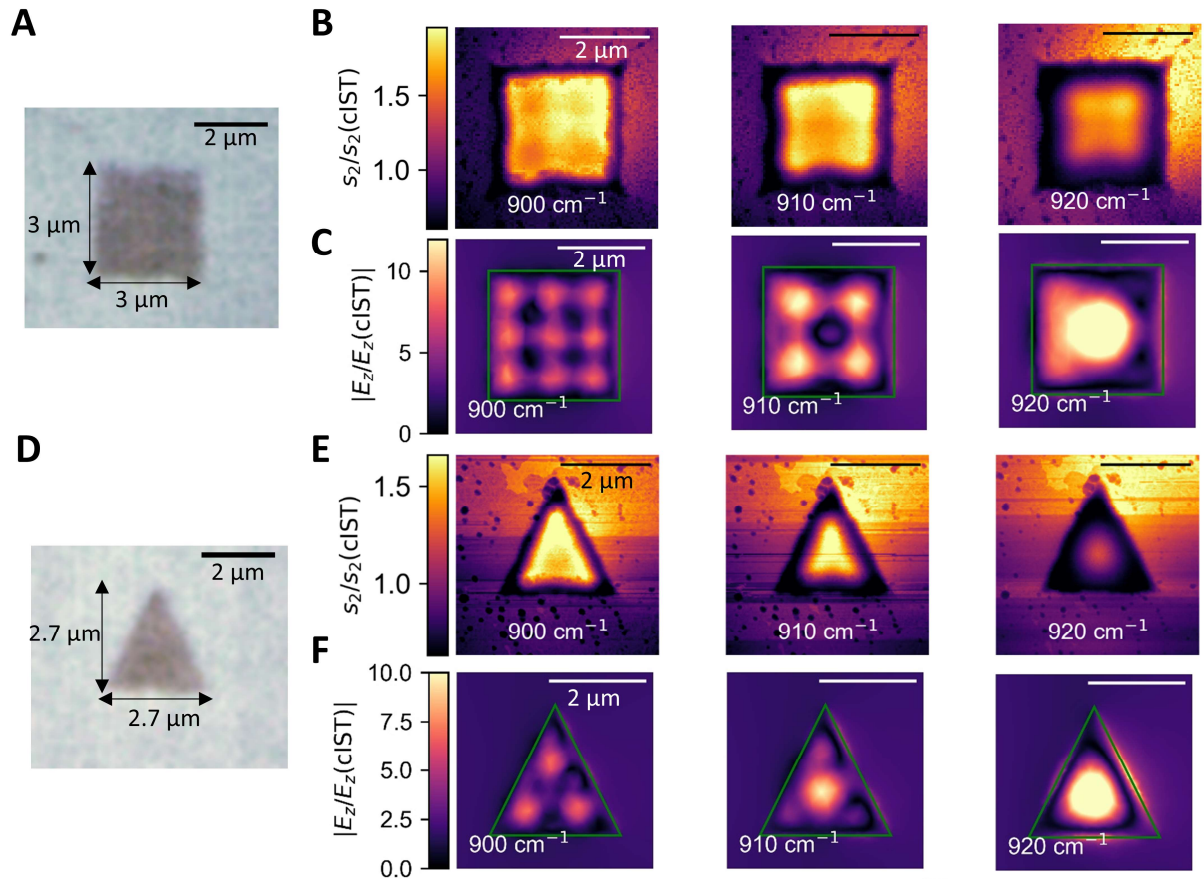

**Figure 5: Investigation of arbitrary resonator shapes. A)** Light microscope image of an amorphous square resonator with a size of 3x3 µm². **B)** Measured near-field amplitude images and numerical field simulations **(C)** of the out-of-plane component of the electric field at different excitation frequencies of 900 cm$^{-1}$, 910 cm$^{-1}$ and 920 cm$^{-1}$. Depending on the excitation frequency, the observed near-field pattern changes drastically. **D)** Light microscope image of a triangle with 2.7 µm edge lengths. **E)** Measured near-field amplitude images and numerical simulations **(F)** at the same excitation frequencies of (B). The scale bars equal 2 µm and the green shapes outline the simulated geometry.

**Discussion:**

In summary, we have demonstrated the potential of a new platform for reconfigurable polariton photonics on the polar crystal SiC employing the plasmonic phase-change material IST. Confined SPhP resonators are directly written and reconfigured inside the material to tune the mode confinement. The vast flexibility of our concept is shown by programming square and triangular resonators without cumbersome fabrication techniques such as electron beam lithography or focused ion beam milling.

Since the plasmonic PCM IST is metallic in the entire infrared spectral range, the presented concept can be easily transferred to other polariton hosting materials. For example, surface plasmon polaritons could be investigated in highly doped semiconductors such as CdO[32] with new widely-tunable laser sources[33] or hyperbolic phonon polaritons of 2d materials such as hexagonal boron nitride (hBN)[34] can be exploited with laser written IST structures. Here, a larger propagation length and therefore the observation of higher order modes could be achieved by combining the plasmonic PCM IST with isotopically pure hBN facilitating less losses.[35–37] We performed numerical simulations of amorphous IST cavities below 100 nm of hBN (see **Supplementary Note 9**) featuring enhanced field pattens inside the resonator and hyperbolic phonon polaritons launched by the crystalline edges outside of the resonators. Consequently, prestructuring the hBN into resonators is not required anymore, facilitating a convenient way for programmable polariton nanophotonics.

Even smaller cavities can be achieved by applying shorter wavelengths for switching the PCM (UV-lasers[38]), by increasing the numerical aperture or by exploiting more sophisticated spatially overlapping pulses, resulting in deep-subwavelength cavities with even more confined fields. Recently, Sheinfux et al. utilized strongly confined polariton resonators in hBN on nanopatterned holes inside a gold film in combination with bound state in continuum (BIC) interference to achieve tremendous quality factors up to 400.[39] This approach can be directly transferred to our demonstrated method, even allowing for reprogramming those cavities and fine-tuning of the investigated field confinements.

Another promising approach might be the programming of anisotropic SPhP resonators by exploiting intrinsically anisotropic materials such as α-$MoO_3$[40,41], $V_2O_5$[42], or β-$Ga_2O_3$[43] featuring a hyperbolic dispersion. These resonators would allow programming of complex non-symmetric resonance modes.

Our work paves the way towards a new powerful platform for mid-IR reconfigurable polaritonic nanophotonics including metasurfaces[44,45], surface-enhanced infrared absorption (SEIRA)[12,46,47] and new sensor technologies with improved sensitivity.

**Experimental Methods:**

**Sample fabrication:**

A layer of amorphous In$_3$SbTe$_2$ and afterwards a capping layer of ZnS:SiO$_2$ were sputtered on a polished 6H-SiC substrate with direct current and radio frequency magnetron sputtering. The thicknesses of the IST layer and the ZnS:SiO$_2$ capping layer are determined to be 35 nm and 15 nm, respectively.

**Optical switching:**

For the laser-induced phase transition, a home-built laser switching setup is used. The light of a laser diode with a wavelength of 660 nm was focused on the sample, which was placed on a piezo stage. For crystallization, 100 single pulses with a power of 80 mW and a pulse length of 3 µs were used to create a crystallized spot. Several spots were placed next to each other to obtain crystallized areas with an amorphous as-deposited region left out in the center. Reamorphization was achieved with a single pulse with a power of 300 mW and a pulse length of 20 ns.

**SNOM:**

Optical characterization was performed with a commercial scattering-type scanning near-field optical microscope by Neaspec GmbH operated in pseudo-heterodyne detection mode.[48] To measure the local electric field, second demodulation order optical near-field amplitude $s_2$ and phase $\varphi_2$ were extracted. Topography was recorded with tapping-mode atomic force microscopy with a PtIr5-coated silicon tip by NanoWorld AG with a curvature radius ≈ 20 nm and resonance frequency ≈ 260 kHz. A tunable quantum cascade laser MIRcat-QT by Daylight Solutions Inc. in continuous wave mode was employed as a light source. The scan included an incident power of 3 mW to 7 mW, a tapping amplitude of about 120 nm (except for about 300 nm for the 3 µm cavity) and a scan speed of about 1.9 µm/s with a scan resolution of 25 nm/pixel.

**Simulations:**

Numerical simulations were performed with CST Studio Suite from Dassault Systèmes. Floquet port excitation with an incident angle of the p-polarized light of 60° normal to the surface and periodic boundary conditions in lateral dimensions were assumed. The electric field calculated with the frequency domain solver was extracted 1 nm above the surface. The s-SNOM tip is not considered in the simulations.

**Supporting Information**:

Supplementary Note 1: Infrared optical properties of $In_3SbTe_2$ (IST)

Supplementary Note 2: Polariton fitting

Supplementary Note 3: SPhP dispersion

Supplementary Note 4: Writing procedure of confined resonators

Supplementary Note 5: SNOM images for different diameters

Supplementary Note 6: Analytical model of circular resonators

Supplementary Note 7: Resonator arrays investigated in the far-field

Supplementary Note 8: Reamorphization of cavities

Supplementary Note 9: Hyperbolic Phonon Polariton cavities with hBN

**Acknowledgements**

L.C., K.W. and T.T. conceived the research idea; L.C., L.S. and K.W designed the research; L.S. carried out the optical switching and the s-SNOM measurements; L.C. and L.S. analyzed the data and carried out the numerical simulations. M.W. provided the sputtering equipment and phase-change material expertise; all authors contributed to writing the manuscript. The authors thank Maike Kreutz for the sputter deposition of the thin film layer stack.
The authors acknowledge support by the Deutsche Forschungsgemeinschaft (DFG No. 518913417 & SFB 917 "Nanoswitches").

**Title:**

**Direct programming of confined Surface Phonon Polariton Resonators with the plasmonic Phase-Change Material In$_3$SbTe$_2$**

*Author(s), and Corresponding Author(s)\**

*Lukas Conrads[+,\*], Luis Schüler[+], Konstantin G. Wirth, Matthias Wuttig, Thomas Taubner[#]*

+ authors contributed equally

**This PDF file includes:**

# Supplementary Note 1: Infrared optical properties of In$_3$SbTe$_2$ (IST)

The dielectric function of IST is obtained with infrared spectroscopy and ellipsometry. The details are taken from Heßler et al.[1]:

In general, we apply the 'Tauc-Lorentz Dispersion formula' by assuming a Lorentz oscillator model multiplied with the Tauc Joint Density of states. The imaginary part is given as follows:

$$\text{Im}(\epsilon_{\text{TL}}) = \frac{A}{\omega} \frac{\omega_0 \gamma (\omega - \omega_g)^2}{(\omega^2 - \omega_0^2)^2 + \gamma^2 \omega^2} \Theta(\omega - \omega_g) \tag{S1.1}$$

Accordingly, $\omega_0$ refers to the oscillator resonance frequency with the damping $\gamma$ and the resonator strength $A$. The band gap frequency is given by $\omega_g$. The real part can be calculated via Kramers-Kronig relation with the effective polarizability $\epsilon_\infty$.

Furthermore, due to the free charge carriers for crystalline IST, an additional Drude term must be added:

$$\epsilon_{\text{Drude}} = -\frac{\omega_p^2}{\omega^2 + i\omega\gamma_D} \tag{S1.2}$$

The plasma frequency is denoted with $\omega_p$, and the damping with $\gamma_D$.

The corresponding parameters of the Tauc-Lorentz-Drude model for amorphous and crystalline IST are shown in **Table S1**. The dielectric function is displayed in **Figure 1A** in the main text.

**Table S1.** Tauc-Lorentz-Drude model parameters for IST.

|  | amorphous IST | crystalline IST |
|---|---|---|
| $A$ [Hz] | 13.0·10$^{16}$ | 4.0·10$^{16}$ |
| $\omega_0$ [Hz] | 4.1·10$^{15}$ | 4.1·10$^{15}$ |
| $\omega_g$ [Hz] | 0.9·10$^{15}$ | 0 |
| $\gamma$ [Hz] | 5.2·10$^{15}$ | 4.1·10$^{15}$ |
| $\omega_p$ [Hz] | - | 7.0·10$^{15}$ |
| $\gamma_D$ [Hz] | - | 0.5·10$^{15}$ |
| $\varepsilon_\infty$ | 2 | 1.4 |

## Supplementary Note 2: Polariton fitting

A sketch of the experimental situation for polariton imaging is shown in **Figure S2.1**. The local electric field probed by the SNOM tip $E_{tip}$ is given by the interference of the illuminating light $E_i$ and the surface phonon polariton field $E_p$: $E_{tip}$ = $E_i$ + $E_p$. Polaritons can be launched at the tip and at the interface between the crystalline and amorphous IST.

Here, three contributions to $E_p$ are considered: The electric field from polaritons launched at the tip $E_{p,t}$, the field from polaritons launched at the boundary $E_{p,b}$, and the field from polaritons launched at the tip and reflected at the interface $E_{p,tb}$. Combining these contributions yields $E_p$ = $E_{p,t}$ + $E_{p,tb}$ + $E_{p,b}$.

Tip-launched polaritons have a contribution $E_{p,t}$ = $\eta_t E_i$, where $\eta_t$ describes the launching efficiency of the tip.[2,3] For the tip-launched polaritons reflected at the boundary, a propagation term has to be added: $E_{p,tb}$ = $\eta_t E_i r_b$ exp($i2k_x(x - x_0)$), where $r_b$ is a reflection coefficient, $k_x$ is the polariton wavevector, $x$ is the position of the tip, and $x_0$ is the position of the boundary.[2,3]

The incident field at the boundary is given by $E_{i,b} = E_i \exp(-ik_i \sin(\alpha)(x - x_0))$, where $k_i = k_{i,0}\sqrt{\epsilon_{aIST}}$, with $k_{i,0}$ = $2\pi\nu$, is the incident wavevector in the amorphous IST, and $\alpha$ is the angle of incidence relative to the sample normal. The field at the tip resulting from the boundary-launched SPhP is then given by $E_{p,b}$ = $\eta_b E_{i,b}$ exp($i(k_x(x - x_0) + k_z z + \varphi_e)$), where $\varphi_e$ is an excitation phase.[4]

The backscattered electric field from the tip at the detector is given by $E = \alpha_{eff}E_{tip}$. The effective polarizability strongly depends on the tip-sample distance, and the tapping amplitude is in general significantly smaller than the extent of the polariton fields in $z$- direction. This means that the influence of the z-dependence of the polariton fields on the detector signal is negligible compared to the z-dependence of $\alpha_{eff}$, and overall, the local field at the tip can be approximated by the field at $z = 0$. Therefore, the $n$th harmonic of the total electric field at the detector is given by[4]

$$E_n = \alpha_{eff,n} E_i \left(1 + \eta_t + \eta_t r_b e^{i2k_x(x-x_0)} + \eta_b e^{i[(k_x - k_i \sin(\alpha))(x-x_0) + \varphi_e]}\right), \quad (S2.1)$$

with $k_i = 2\pi\nu\sqrt{\epsilon_{aIST}}$.

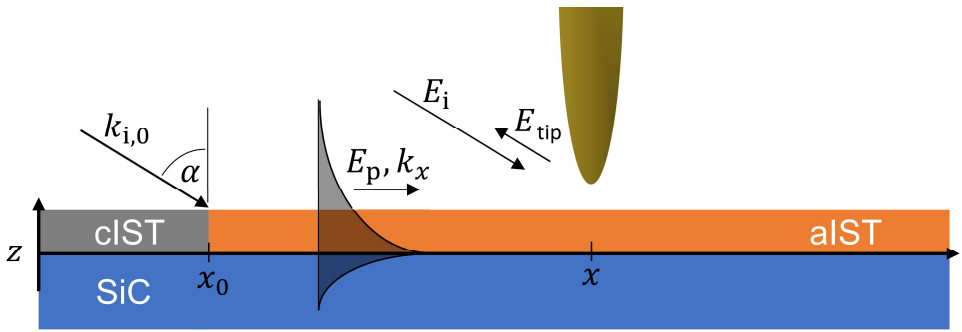

**Figure S2.1:** Sketch of polariton imaging. The incident light ($k_{i,0}, E_i$) illuminates the SNOM tip at position $x$ and the boundary of cIST/aIST at position $x_0$. A surface phonon polariton with wavevector $k_x$ is launched. The electric field $E_{tip}$ is scattered back from the tip.

To estimate the polariton wavevector from a SNOM scan, the $n$th demodulation order optical near-field amplitude $s_n$ is referenced to the amplitude at a position where no polaritons are present, in this case $s_n(cIST)$. The result, $s_n/s_n(cIST)$, is then given by the absolute value of the referenced electric field $|E_n/(\alpha_{eff,n}E_i)|$.[4]

Since the factors $\eta_t$ and $r_b$ in the term $\eta_t r_b$ from (S2.1) cannot be separated during fitting, they are combined into $\eta_{tb} = \eta_t r_b$. A linear background was used to improve the fits, resulting in a function:

$$f = \left|1 + \eta_t + \eta_{tb} e^{i2k_x(x-x_0)} + \eta_b e^{i[(k_x - 2\pi\nu\sqrt{\epsilon_{aIST}}\sin(\alpha))(x-x_0) + \varphi_e]} + b(x - x_0)\right|, \quad (S2.2)$$

with the fitting parameters $k_x$, $\eta_t$, $\eta_{tb}$, $\eta_b$, $b$ and $\varphi_e$.

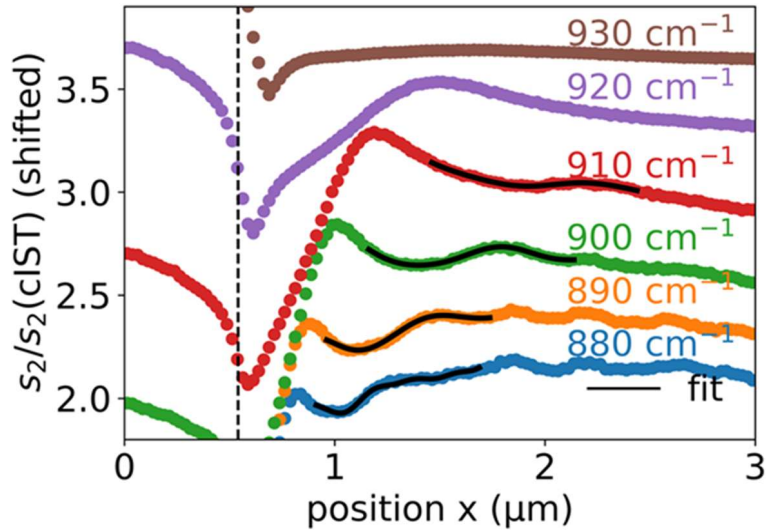

**Figure S2.2:** Zoom-in to the fits presented in **Figure 2C.**

## Supplementary Note 3: SPhP Dispersion

The SPhP dispersion relation for the given layer stack is obtained via two ways. First, the reflection coefficient for p-polarized light $r_\mathrm{p}$ is calculated for different wavevectors and wavenumbers with the transfer matrix method (TMM), in which the electric field in the top layer is related to the field in the bottom layer by matrices describing propagation within the different layers and transmission between them.[5,6] The dispersion relation is then given by the points at which the absolute value of the imaginary part of $r_\mathrm{p}$, $|\mathrm{Im}(r_\mathrm{p})|$, diverges.[7]

The other way, which only works for three layers, is to derive the reflection coefficient for a (thin) homogeneous dielectric film and solve the implicit equation:

$$e^{-2k_{2,z}d} = \frac{\frac{k_{1,z}}{\epsilon_1}+\frac{k_{2,z}}{\epsilon_2}}{\frac{k_{1,z}}{\epsilon_1}-\frac{k_{2,z}}{\epsilon_2}} \cdot \frac{\frac{k_{3,z}}{\epsilon_3}+\frac{k_{2,z}}{\epsilon_2}}{\frac{k_{3,z}}{\epsilon_3}-\frac{k_{2,z}}{\epsilon_2}}, \quad (S3.1)$$

which describes the points at which the reflection coefficient diverges.[7–9] Here, $\epsilon_l$ is the effective permittivity in layer $l$, $d$ is the thickness of the second layer (the first and third layer are assumed to be infinitely thick), and $k_{l,z}^2 = k_\mathrm{sp}^2 - k_0^2\epsilon_l$, where $k_\mathrm{sp}$ is the SPhP wavevector and $k_0$ is the wavevector of the incident light in vacuum. Since the effective permittivity is a function of the frequency ν, this equation can be solved numerically, yielding a dispersion relation $\nu(k_\mathrm{sp})$ or $k_\mathrm{sp}(\nu)$.

There are three relevant solutions:[2] For the case of a localized SPhP, for example confined in a resonator, $k_\mathrm{sp}$ can be determined by the resonance condition and solving eq. S3.1 numerically for $\nu(k_\mathrm{sp})$ yields the resonant frequency. If propagating SPhPs are considered, the frequency of the SPhP is defined by the exciting light[2]. Solving for $k_\mathrm{sp}$ yields two results, called slow and fast SPhP, respectively[10]. The wavevectors of the fast SPhP are close to the light line $|k_\mathrm{sp}| \approx |k_0\sqrt{\epsilon_l}|$ while the slow SPhP can have very large wavevectors with $|k_\mathrm{sp}| \gg |k_0\sqrt{\epsilon_l}|$.

In **Figure S3**, the solutions corresponding to the slow (green line) and fast (brown line) SPhP (solved for $k_\mathrm{sp}(\nu)$) and the solution corresponding to a localized (blue line) SPhP (solved for $\nu(k_\mathrm{sp})$) for a three-layer case of SiC/aIST/ZnS:SiO$_2$ and the absolute value of Im($r_\mathrm{p}$) calculated with TMM for a layer stack of SiC/aIST/ZnS:SiO$_2$/air are plotted. The higher effective permittivity of the capping layer compared to air shifts the dispersion to lower frequencies.

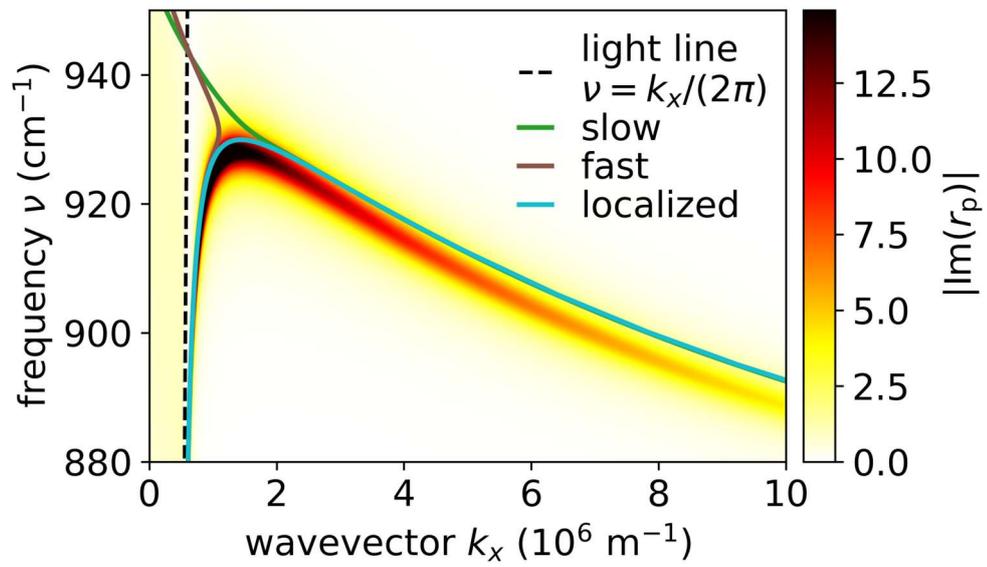

**Figure S3:** The three solutions of the dispersion relation for a three-layer case (SiC/aIST/air) and the dispersion calculated with TMM for the layer stack SiC/aIST/ZnS:SiO$_2$/air. The solution of localized SPhPs equals the fast (slow) SPhP for small (large) wavevectors.

## Supplementary Note 4: Writing procedure of confined resonators

Confined SPhP resonators are directly optically written inside the amorphous IST layer with precisely arranged laser pulses. The schematic working principle is shown in **Figure S4**. First, multiple crystallization spots are positioned next to each other with vertical distance Δy = 1 µm and horizontal distance Δx = 1.3 µm (i). The larger horizontal distance is caused by the intrinsic elliptical shape of the laser beam. Afterwards, 25 spots circularly arranged are added, resulting in the shown circular cavities from Figure 3 (ii). Finally, these cavities are locally addressed again, and the diameter is reduced by applying again 25 spots within a smaller circle (iii). All spots overlap and hence result in a homogeneous crystalline area with cavities of amorphous IST in the center.

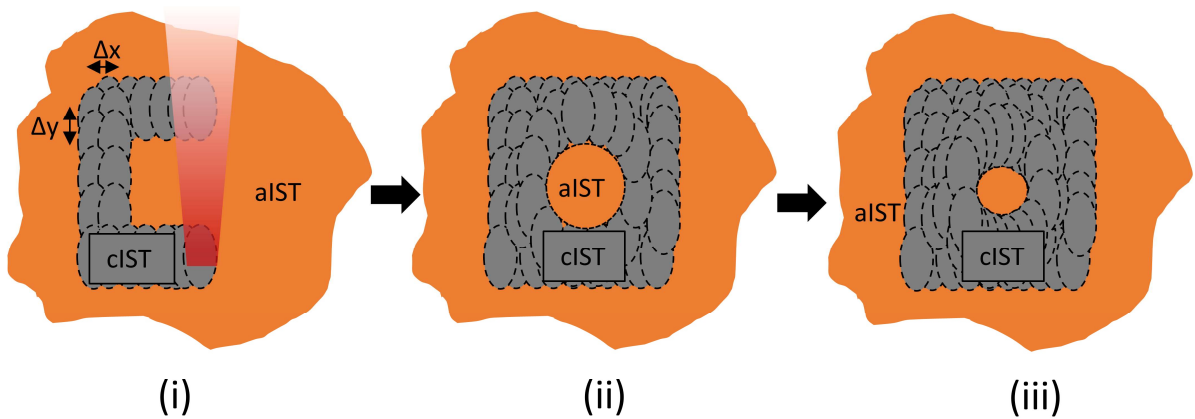

**Figure S4:** Schematic writing process of confined resonators.

# Supplementary Note 5: Analytical model of circular resonators

For a circular resonator or cavity with diameter $D$, the resonance condition for a mode $(m, n)$ is given by

$$k_{\text{sp}}(\nu)D + \phi = 2x_n(J_m), \tag{S4.1}$$

where $\phi$ is the phase increment caused by the reflection at the cavity boundary and $x_n(J_m)$ is the $n$th zero of the Bessel function of the first kind $J_m$.[11]

For a given cavity diameter, $k_{\text{sp}}(\nu)$ can be calculated for different modes with eq. S4.1. Afterwards, the resonant frequency is determined by the point where $|\text{Im}(r_p)|$, calculated with the TMM for four layers (air/capping/IST/SiC), is maximal for the given $k_{\text{sp}}(\nu)$.

The resonant frequency in dependence of the cavity diameter for the (0, 1) mode and two reflection phases is shown in **Figure S5**, together with the experimental data. The experimental resonance positions were estimated from the SNOM images (**Figure 3** and **Figure S6**) and fit to the analytical model if a reflection phase $\phi = -\pi$ is assumed.

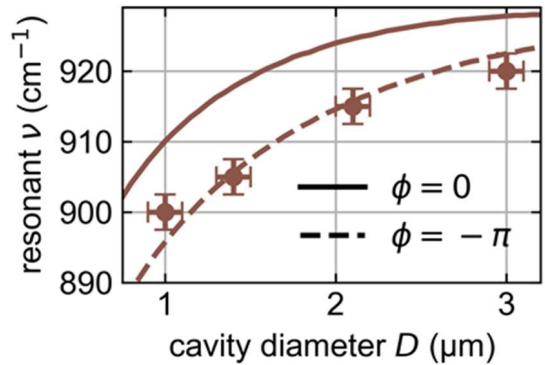

**Figure S5:** Analytical relationship between cavity diameter and resonant frequency for different reflection phases and experimental data.

**Supplementary Note 6: SNOM images of cavities with different diameters**

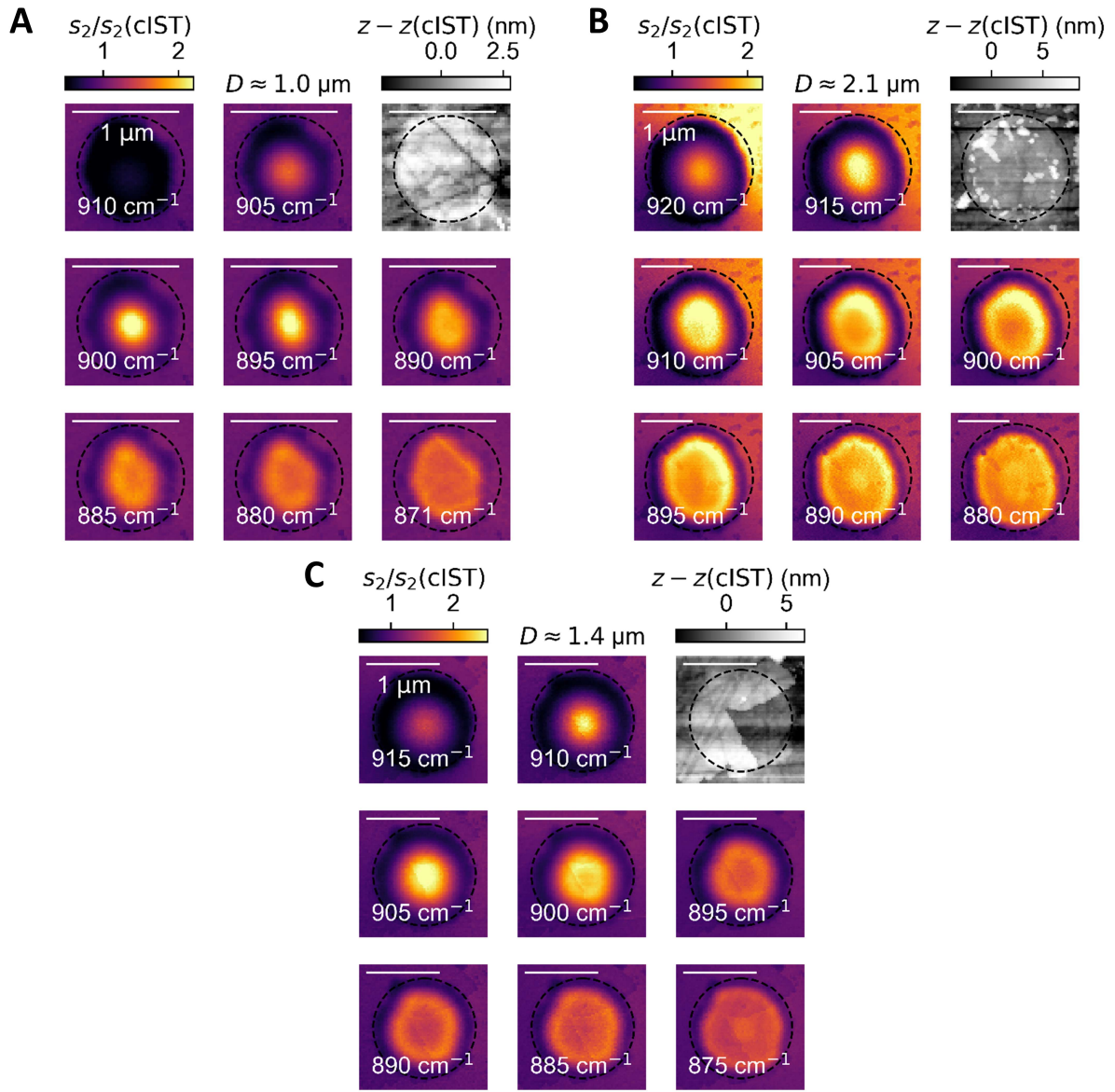

**Figure S6.** SNOM images and topography of cavities with diameters of about 1 μm, 2.1 μm and 1.4 μm. The circles indicate the cavity boundary. Due to dirt and other surface contaminations of the ZnS:SiO$_2$, the visibility of the cavities in the AFM images is low.

**Supplementary Note 7: Resonator arrays investigated in the far-field**

For four selected diameters, multiple cavities with the same diameter were arranged in an array with an area of about 40 x 40 µm$^2$. Optical images of these arrays are shown in **Figure S7A**. The arrays were measured with a Fourier-transform infrared microscope in a grazing incidence reflection setup, with incidence angles ranging from 52° to 84°.[12]

The amorphous surrounding was blocked out with knife-edge apertures. **Figure S7B** displays the measured spectra, which are shifted for better visibility and were referenced to an area of crystallized IST. In each spectrum, a dip in reflectance (marked by colored arrows) is visible, caused by the SPhP resonance in the cavity. The near-field resonance position does not coincide with the far-field one. This behavior is known from theory and experiments.[13]

The spectra in **Figure S7C** were simulated with CST, under the assumption of an incidence angle of 70°. There, smaller dips at lower wavenumbers are present, corresponding to higher order modes. Such dips can hardly be discerned in the experimental spectra. In general, a good agreement between simulated and the experimentally observed spectra is present. The lower magnitude and increased broadness of the experimental peaks can be attributed to the broad angular distribution of the objective used and imperfect varying resonators across the array, especially a varying diameter and no perfectly round borders due to the stochastic nature of the crystallization.

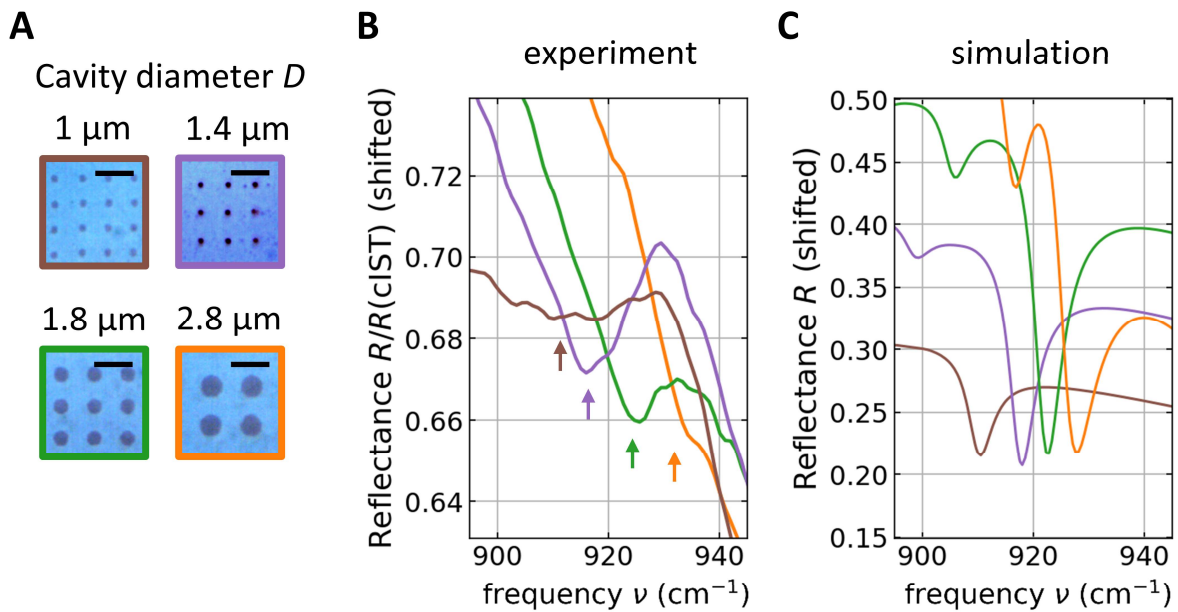

**Figure S7**. Far-field measurements and simulations of cavity arrays. **A)** Optical microscope images of a section of the arrays of resonators with different diameters. The scale bars correspond to 5 µm. **B), C)** Experimental **(B)** and simulated **(C)** reflectance. The arrows are a guide to the eye and indicate dips in the reflectance corresponding to the cavity resonance.

## Supplementary Note 8: Reamorphization of cavities

Previously, the cavities have been programmed by crystallizing the amorphous IST around the cavity (see **Figure S8A**). To demonstrate the reversibility of the optical writing, cavities were created by crystallizing the whole area and then reamorphizing a circular area in the center. This process is sketched in **Figure S8B**. In **Figure S8D**, SNOM images of such a cavity are compared to the images of cavity with a diameter $D \approx 1.4$ µm (**Figure S8C**), fabricated with the initial method (c.f. **Figure S8C**). The SNOM images look similar, but deformation of the sample (see **Figure S8E**) and resulting higher average distance to the SiC/IST interface, where the polariton field is highest, leads to reduced signal and visibility. A possible explanation for the deformation is that the IST expands rapidly when heated, thus deforming and possibly cracking the capping.[14] At 880 cm$^{-1}$, the ring-like feature with higher signal cannot be discerned anymore. An optimization of the laser parameters might lead to decreased deformation and therefore increased reversibility.

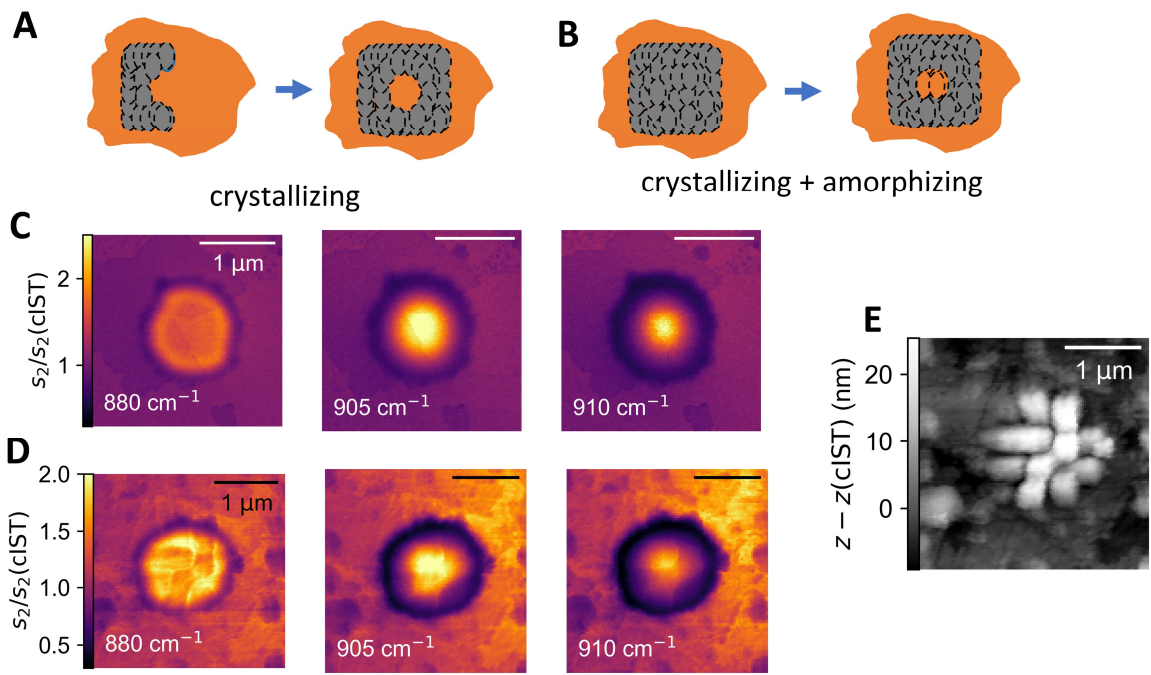

**Figure S8.** Reamorphized cavities. **A), B)** Sketch of the optical writing process used to define the cavities with the help of crystallizing (**A**) or crystallizing and amorphizing (**B**) pulses. The aIST is colored red, cIST blue. **C), D)** SNOM images at different frequencies of the cavities created without (**C**) and with (**D**) amorphizing pulses. **E)** Topography of the reamorphized cavity.

## Supplementary Note 9: Hyperbolic Phonon Polariton cavities with hBN

We demonstrated our concept of direct programming SPhP resonators inside IST on the polar crystal SiC. However, the concept can be easily transferred to other material systems such as 2d hyperbolic materials, e.g., hexagonal boron nitride (hBN). Therefore, we performed numerical simulations of an amorphous IST cavity with a diameter $D$ = 750 nm buried below a 100 nm thin hBN layer (see **Figure S9A**). The dielectric function of hBN with the lower and upper Reststrahlenband is shown in **Figure S9B**. The upper Reststrahlenband is located between 1360-1610 cm$^{-1}$ with Re($\varepsilon_z$) > 0 and Re($\varepsilon_t$) < 0 showing a hyperbolic response. Normalized electric field simulations of the E$_z$-component for three different excitation frequencies are displayed in **Figure S9C**. Inside the cavities similar mode patterns compared to the IST resonators on SiC can be observed. Outside of the resonators, concentric rings corresponding to propagating polaritons are visible.

Hence, the concept of programming confined resonators inside the PCM is not limited to the polar crystal SiC but can be easily transferred to other materials such as hBN or anisotropic crystals.[15–17]

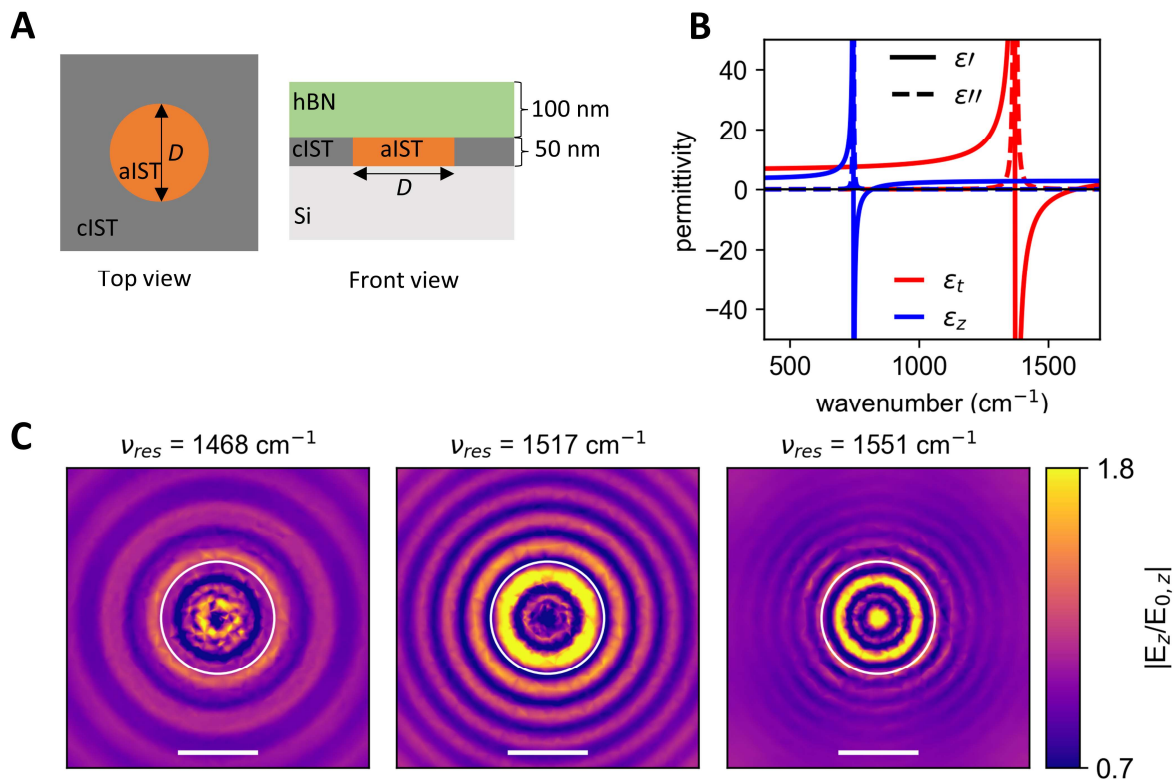

**Figure S9: Simulations of IST cavities below an hBN flake. A)** Schematic sketch of the simulated layerstack with an amorphous circular cavity with a diameter $D$ = 750 nm buried below the 100 nm hBN. **B)** Dielectric function of the in-plane $\varepsilon_t$ and out-of-plane $\varepsilon_z$ component.

**C)** Simulated out-of plane component of the electric field normalized to the incident field 1 nm above the hBN for frequencies in the upper Reststrahlenband. Polariton fringes are clearly visible, as well as different modes inside the cavity. The scale bars equal 500 nm.